\documentclass[twocolumn]{autart}    

\usepackage{amssymb}
\usepackage[dvips]{epsfig}
\usepackage{amsmath,amssymb}
\usepackage{latexsym}
\usepackage{subfigure}
\usepackage{color}
\usepackage[center]{caption}
\usepackage{graphicx}
\usepackage{url}

\newcommand{\be}{\begin{equation}}
\newcommand{\eq}{\end{equation}}
\newcommand{\ba}{\begin{array}}
\newcommand{\ea}{\end{array}}
\newcommand{\bean}{\begin{eqnarray*}}
\newcommand{\eean}{\end{eqnarray*}}
\newcommand{\bea}{\begin{eqnarray}}
\newcommand{\eea}{\end{eqnarray}}

\newcommand{\R}{\rm {I\kern-2pt R}}

\newcommand{\beq}{\begin{equation}}
\newcommand{\eeq}{\end{equation}}

\newtheorem{theorem}{\bf Theorem}[section]
\newtheorem{remark}{\bf Remark}[section]
\newtheorem{lemma}{\bf Lemma}[section]

\newtheorem{proposition}{\bf Proposition}[section]
\newtheorem{definition}{\bf Definition}[section]

\begin{document}

\begin{frontmatter}

\title{Exact exponential synchronization rate of high-dimensional Kuramoto models with identical oscillators and digraphs} 
\thanks[footnoteinfo]{This work is supported in part by National Natural Science Foundation
(NNSF) of China under Grants 61673012 and 11971240.
}

\author[NNU,SEU]{Shanshan Peng}\ead{shanshan\b{~}peng@hotmail.com},
\author[SDNU]{Jinxing Zhang}\ead{jxingzhang@hotmail.com},  
\author[NNU]{Jiandong Zhu}\ead{zhujiandong@njnu.edu.cn},
\author[SEU]{Jianquan Lu}\ead{jqluma@seu.edu.cn},
\author[SDNU]{Xiaodi Li}\ead{lxd@sdnu.edu.cn}


\address[NNU]{School of Mathematical Sciences, Nanjing Normal University, Nanjing,
        210023, PRC}  
\address[SEU]{School of Mathematics, Southeast University, Nanjing, 211189, PRC}
\address[SDNU]{School of Mathematics and Statistics, Shandong Normal University, Jinan, 250358, PRC}

\begin{keyword}                           
High-dimensional Kuramoto model; Exponential synchronization rate; Directed graph.               
\end{keyword}                             

\begin{abstract}
For the high-dimensional Kuramoto model with identical oscillators under a general digraph that has a directed spanning tree, although exponential synchronization was proved under some initial state constraints, the exact exponential synchronization rate has not been revealed until now. In this paper, the exponential synchronization rate is precisely determined as the smallest non-zero real part of Laplacian eigenvalues of the digraph. 
Our obtained result extends the existing results from the special case of strongly connected balanced digraphs to the condition of general digraphs owning directed spanning trees, which is the weakest condition for synchronization from the aspect of network structure. Moreover, our adopted method is completely different from and much more elementary than the previous differential geometry method.
%
\end{abstract}
\end{frontmatter}

\section{Introduction}
The well-known Kuramoto model proposed by Yoshiki Kuramoto in \cite{Kuramoto1975Self-entrainment} is one of the most successful mathematical models to describe collective behaviors of complex dynamical networks. Kuramoto model and its various generalized forms have many applications in engineering, neuroscience, physics and so on \cite{RODRIGUES20161}.  For Kuramoto model, exponential synchronization is an typical collective behavior and an interesting theoretical issue \cite{Chopra2009On}, \cite{Dong2013Synchronization},  \cite{TONG2018129}, \cite{Wang2013Exponential}. The dynamics of the general Kuramoto model is described by
\begin{equation}
\label{eq1}
\begin{split}
\dot \theta_{i}&=\omega_{i}+k\sum\limits_{j=1}^ma_{ij}\sin(\theta_{j}-\theta_{i}),\ \ i=1,2,\dots,m,
\end{split}
\end{equation}
where the $m\times m$ matrix $(a_{ij})$ is the adjacency matrix of the interconnecting graph, $\theta_i$ is the $i$-th oscillator's phase angle and  $\omega_i$ is the natural frequency.
The $n$-dimensional vector-valued Kuramoto
model is described as follows:
\begin{equation}
 \label{eq2}
  \dot{r}_{i}=\Omega_{i}r_{i}+k\sum\limits_{j=1}^{m}a_{ij}(r_{j}-\frac{r_{i}^{T}r_{j}}{r_{i}^{T}r_{i}}r_{i}),\ i=1,2,\dots,m,
\end{equation}
where $r_{i}\in{\mathbf{R}^{n}}$ is the $i$-th oscillator's state,
$\Omega_{i}$ is an $n\times n$ skew-symmetric matrix. When $n=2$ and $r_i=[\cos \theta_i,\ \sin\theta_i]^\mathrm{T}$, by the $n$-dimensional Kuramoto model (\ref{eq2}), it is easy to deduce the original Kuramoto model (\ref{eq1}). If (\ref{eq2}) has the identical oscillators, i.e. $\Omega_i=\Omega$ for $i=1,2,\dots,m$, one can assume, without loss of generality, that   $\Omega_i=0$ for $i=1,2,\dots,m$ \cite{ZHU2013Synchronization}.
Limited on the unit sphere $\mathbf{S}^{n-1}$, dynamical network (\ref{eq2}) with $\Omega_i=0$ and $k=1$
is reduced to
\begin{equation}
 \label{eq3}
  \dot{r}_{i}=\sum\limits_{j=1}^{m}a_{ij}\big(r_{j}-(r_{i}^{T}r_{j})r_{i}\big),\ i=1,2,\dots,m,
\end{equation}
which is first proposed in \cite{Olfati-Saber2006Swarms}
as a swarm model on spheres, and can be used to solve the max-cut
problem. Due to Lohe's pioneering literatures such as \cite{Lohe2009Non} and
\cite{Lohe2010Quantum}, a class of high-dimensional Kuramoto models is also called Lohe model, which has potential applications to quantum systems. More general Kuramoto models defined on some matrix manifolds can be seen in \cite{Jared2020A},   \cite{MARKDAHL2020High},  \cite{Sarlette2009Consensus}.

For the case of complete graphs, many theoretical results on the synchronization of high-dimensional Kuramoto model have been achieved such as  \cite{choi2014} , \cite{Olfati-Saber2006Swarms} and \cite{Li2014Unified}.
In our early paper \cite{ZHU2013Synchronization}, the synchronization limited on an open half-sphere is proved for the case of general undirected connected graphs. If the states of particles are not limited on an open half-sphere, some almost global synchronization results can also be obtained for the case of undirected connected graphs \cite{Markdahl2018Almost}, \cite{zhu2014high}.

For general directed graphs, the synchronization of the high-dimensional Kuramoto model is proved in \cite{Lageman2016Consensus}
by using a differential geometry method. However, in our paper \cite{Zhang2018OnEquilibria}, a simpler approach based on LaSalle invariance principle is adopted to demonstrate the synchronization under the general directed graph condition. Just like the original Kuramoto model, high-dimensional Kuramoto model also has the dynamical property of exponential synchronization (see \cite{Chi2014Emergent}, \cite{choi2014}, \cite{Lageman2016Consensus} and \cite{ZHANG2019Exponential}). But so far, the exact exponential synchronization rate is only obtained for a very special kind of digraphs, i.e. strongly connected balanced digraphs.
In our recent paper \cite{ZHANG2019Exponential}, the exponential synchronization is achieved for a general digraph admitting a spanning tree, but the exact exponential synchronization rate has not yet been obtained.

In this paper, for the
high-dimensional Kuramoto model under a general digraph containing
a spanning tree, it is proved the exponential synchronization rate is also exactly $\mathrm{Re}(\lambda_2)$ just like the case of strongly connected balanced digraphs considered in \cite{Lageman2016Consensus}, where $\lambda_2$ is the eigenvalue of the Laplacian $L$ with the smallest non-zero real
part.

 \par
 The rest of this paper is organized as follows. Section 2 gives some preliminaries and the
 problem statement.
 Section 3 includes our main results. Section 4 is a summary.\par

\section{Preliminaries and Problem Statement}
For high-dimensional Kuramoto model (\ref{eq3}), exponential synchronization is defined as follows:
\begin{definition}
\label{def2.1}
It is said that the {\it
exponential synchronization} for  (\ref{eq3}) is achieved if there
exists a $\mu>0$ and a constant $c>0$  such that
\begin{equation}
\label{eq4}
|| r_i(t)-r_j(t)||\leq c  {\rm e}^{-\mu t} \ \ \forall  \ i,\  j=1,2,\dots,m.
\end{equation}
Moreover, we say that the exponential synchronization rate is at least $\mu$ when (\ref{eq4}) holds. The minimum $\mu$ satisfying (\ref{eq4}) is just called the exponential synchronization rate.
\end{definition}

The first result on the exponential synchronization of $(\ref{eq3})$ under
digraphs is obtained in \cite{Lageman2016Consensus}, which is
based on differential geometry method. The main contribution of Theorem 1 and  Corollary 1 of  \cite{Lageman2016Consensus} can be rewritten as follows:
\begin{proposition}
\label{prop1} Consider the high-dimensional Kuramoto model $(\ref{eq3})$.
If the interconnecting graph is strongly connected and balanced, then the local exponential synchronization is achieved and the exact exponential synchronization rate is the smallest non-zero real
part of the Laplacian eigenvalues of the interconnecting digraph.
\end{proposition}

The target of this paper is to get the exact exponential synchronization rate under the more general digraph condition and the framework of matrix Riccati differential equation proposed in our early paper \cite{ZHANG2019Exponential}.

 Let
\begin{equation}
\label{eq5}
e_{ij}=1-r_{i}^{T}r_{j}=\frac{1}{2}||r_i-r_j||^2.
\end{equation}
It is easily seen that $e_{ij}=e_{ji}$, $e_{ii}=0$ and
$0\leq{e_{ij}}\leq{2}$ for every $i,j=1,2,\dots,m$. Let $E(t)\!=\!(e_{ij}(t))\!\in{\!\mathbf{R}^{m\times{m}}}$. By \cite{ZHANG2019Exponential},
the dynamics of $E(t)$ is described by
matrix Riccati differential equation
\begin{eqnarray}
\label{eq6}
\dot{E}&\!=&\!-LE\!-\!EL^{\!\mathrm T}\!\!- \!\alpha(E)\mathbf{1}^{\!\mathrm T}\!\!- \!\mathbf{1}\alpha^{\!\mathrm T}\!(E)
\!+ \!\Lambda (E)\!E \!+\! E\Lambda (E),
\end{eqnarray}
where $L$ is the Laplacian matrix of the digraph,
\begin{equation}
\label{eq7}
 \alpha(E)\!=\!(\alpha_{1}(E),\alpha_{2}(E),\dots,\alpha_{m}(E))^{\mathrm T}\!\in{\!\mathbf{R}^{m}},
\end{equation}
with each $\alpha_{i}(E)\!=\!\sum\limits_{l=1}^{m}a_{il}e_{il}$, and $\Lambda(E)\!=\!\mathrm{diag(}\alpha(E))$ is the diagonal matrix with the diagonal elements composed of $\alpha_1(E)$, $\alpha_2(E)$, $\dots$, $\alpha_m(E)$.

For the linear space $\mathbf{R}^{m\times m}$, denote by $S_m^0$ the subspace composed of all the symmetric real matrices with all the diagonal entries being $0$. Then the synchronization error equation (\ref{eq6}) is the dynamics restricted on $S_m^0$.
Let $S_m$ and $K_m$ be the $m$-order symmetric matrix subspace and the $m$-order skew-symmetric matrix subspace, respectively. Then
\begin{equation}
\label{eq8}
\mathbf{R}^m=S_m\oplus K_m,\ \ S_m=S_m^0\oplus D_m,
\end{equation}
where $D_m$ is the $m$-order diagonal matrix subspace and $\oplus$ denotes the direct sum.

\section{Main Results}
In our early paper \cite{ZHANG2019Exponential}, Lyapunov functions are designed by using the left eigenvector of the Laplacian matrix of the digraph and consequently the exponential synchronization is proved. However, we cannot get the exact exponential synchronization rate by the approach of Lyapunov functions.

In this section, we turn to the Lyapunov's first method, i.e. the approximate exponential linearization method to get the exact synchronization rate. It is straightforward to check that the approximate linearized system of (\ref{eq6}) can be rewritten as
\begin{eqnarray}
\label{eq9}
\dot{E}&\!=&\!-\big(\!LE\!+\!EL^{\!\mathrm T}\!-\hat L\mathrm{vec}(E)\mathbf{1}_m^{\!\mathrm T}\!- \!\mathbf{1}_m(\mathrm{vec}(E))^{\!\mathrm T}\hat L^{\!\mathrm T}
\!\big),
\end{eqnarray}
where $\mathrm{vec}(\cdot): \mathbf{R}^{m\times m}\rightarrow \mathbf{R}^{m^2}$ denotes the column-stacking operator,
$$
\hat L\!=\!\left[\!\begin{array}{cccc}
          L_1& \\
          & L_2\\
          &&\ddots\\
          &&&L_m
          \end{array}
          \!\right]\!,\ \ L_i\!=\!\mathrm{Row}_i(L),\ \ i\!=\!1,2,\dots,m.
$$
%
%

Considering  (\ref{eq9}), we define a linear transformation on $\mathbf{R}^{m\times m}$ by
\begin{equation}
\label{eq10}
\mathcal{T}(X)=LX\!+\!XL^{\!\mathrm T}\!-\hat L\mathrm{vec}(X)\mathbf{1}_m^{\!\mathrm T}\!- \!\mathbf{1}_m(\mathrm{vec}(X))^{\!\mathrm T}\hat L^{\!\mathrm T}.
\end{equation}
So the exponential decay rate of $E(t)$
is just the smallest non-zero real
part of eigenvalues of $\mathcal{T}(X)$ restricted on $S_m^0$.
It is not easy to directly compute the eigenvalues of $\mathcal{T}(X)$ restricted on $S_m^0$. We first investigate properties of $\mathcal{T}(X)$ on $\mathbf{R}^m$.

\begin{proposition}
\label{prop1}
Consider the linear transformation $\mathcal{T}(X)$ defined by (\ref{eq10}) and the Lyapunov mapping $\mathcal{S}(X)=LX+XL^\mathrm{T}$. For the subspaces $S_m^0$, $D_m$ and $K_m$, the following statements hold:\par
{\rm (i)} if $X\in S_m$, then $\mathcal{T}(X)\in S_m^0$;\par
{\rm (ii)} if $X\in K_m$, then the projection of $\mathcal{T}(X)$ onto $K_m$ is just   $\mathcal{S}(X)$.
\end{proposition}
\begin{pf}
(i) If $X=X^\mathrm{T}$, it is easy to check that $
 \big(\mathcal{T}(X)\big)^\mathrm{T}=\mathcal{T}(X)$.
Moreover, a straightforward computation shows that
\begin{eqnarray}
\delta_i^\mathrm{T}\mathcal{T}(X)\delta_i&=& 2\big(\delta_i^\mathrm{T}LX\delta_i-\delta_i^\mathrm{T}\hat L\mathrm{vec}(X)\mathbf{1}_m^\mathrm{T}\delta_i\big)\nonumber \\ &=&2\big(L_iX_i-\delta_i^\mathrm{T}\hat L\mathrm{vec}(X)\big)=0,\nonumber
\end{eqnarray}
where $\delta_i$ is the $m$-dimensional vector whose $i$-th entry is one and the others are zero. So $\mathcal{T}(X)\in S_m^0$.

(ii) If $X=-X^\mathrm{T}$, then  $$(\mathcal{S}(X))^{\!\mathrm T}=X^{\!\mathrm T}L^{\!\mathrm T}+LX^{\!\mathrm T}=-\mathcal{S}(X).$$
Thus $\mathcal{S}(X)\in K_m$.
From (\ref{eq10}), it follows that
 $$\mathcal{T}(X)=\mathcal{S}(X)-
(\hat L\mathrm{vec}(X)\mathbf{1}_m^{\!\mathrm T}\!+ \!\mathbf{1}_m(\mathrm{vec}(X))^{\!\mathrm T}\hat L^{\!\mathrm T}),
$$
where $\mathcal{S}(\!X\!)\!\in \!K_m$ and $\hat L\mathrm{vec}(\!X\!)\mathbf{1}_m^{\!\mathrm T}\!+ \!\mathbf{1}_m(\!\mathrm{vec}(\!X\!)\!)^{\!\mathrm T}\!\hat L^{\!\mathrm T}\!\in S_m$.
So the projection of $\mathcal{T}(X)$ onto $K_m$ is just  $\mathcal{S}(X)$.\hfill $\blacksquare$
\end{pf}

Denote by $B_1$, $B_2$ and $B_3$ the base of $S_m^0$, $D_m$ and $K_m$, respectively. By conclusion (i) of Proposition \ref{prop1}, we have $\mathcal{T}(S_m^0)\subset S_m^0$ and $\mathcal{T}(D_m)\subset S_m^0$. So
\begin{equation}
\label{eq11}
\mathcal{T}[B_1,\ B_2,\ B_3]=[B_1,\ B_2,\ B_3]
           \left[\begin{array}{ccc}
           T_{11} & T_{12}  & T_{13}\\
                 & 0 &  T_{23}\\
                           && T_{33} \end{array}
           \right].
\end{equation}
By the properties of the Lyapunov mapping $\mathcal{S}(\cdot)$ \cite{Cheng2001On}, \cite{CHENG19993724}, we get
\begin{equation}
\label{eq12}
\mathcal{S}[B_1,\ B_2,\ B_3]=[B_1,\ B_2,\ B_3]
           \left[\begin{array}{ccc}
           S_{11} & S_{12}  & 0\\
           S_{21} & S_{22}  & 0\\
                   &&     S_{33} \end{array}
           \right].
\end{equation}
From conclusion (ii) of Proposition \ref{prop1}, it follows that  $T_{33}=S_{33}$.
Fortunately, the eigenvalues of $S_{33}$, i.e. the spectrum of the Lyapunov mapping $\mathcal{S}(\cdot)$ restricted on $K_m$, have been revealed.
\begin{lemma}\cite{CHENG19993724}
\label{lemma1}
Assume that the eigenvalues of $L$ be $\lambda_1$, $\lambda_2$, $\lambda_3$, $\dots$, $\lambda_m$. Then
the eigenvalues of the Lyapunov mapping $\mathcal{S}(\cdot)$ restricted on $K_m$ are $\{\lambda_i+\lambda_j\ |\ 1\leq i < j\leq m\}$.
\end{lemma}

Therefore, if all the eigenvalues of $\mathcal{T}(\cdot)$ on $\mathbf{R}^{m\times m}$ are obtained, then by (\ref{eq11}) and Lemma \ref{lemma1} we can get the eigenvalues
of $T_{11}$, i.e. the eigenvalues of $\mathcal{T}(\cdot)$ restricted on $S_m^0$.
So, the rest content of this paper is just pursing the eigenvalues of $\mathcal{T}(\cdot)$ on $\mathbf{R}^{m\times m}$.
To this end, we use the isomorphic mapping vec($X$): $\mathbf{R}^{m\times m}\rightarrow \mathbf{R}^{m^2}$ to deal with $\mathcal{T}(\cdot)$.

\begin{lemma}\cite{Magnus1988Matrix}
\label{lemma2}
Let $A$, $B$  and $C$ be $m\times n$, $n\times s$ and $s\times t$ matrices, respectively. Then
$$\mathrm{vec}(ABC) = (C^\mathrm{T} \otimes A)\mathrm{vec}(B).$$
\end{lemma}

Applying Lemma \ref{lemma2} to $\mathcal{T}(X)$ described by (\ref{eq10}) yields
\begin{equation}
\mathrm{vec}(\mathcal{T}(X))\!=\!(L\otimes I_m+I_m\otimes L-\mathbf{1}_m\otimes \hat L-\hat L\otimes \mathbf{1}_m)\mathrm{vec}(X).\nonumber
\end{equation}
So the eigenvalues are just those of the matrix
\begin{equation}
\label{eq13}
T=L\otimes I_m+I_m\otimes L-\mathbf{1}_m\otimes \hat L-\hat L\otimes \mathbf{1}_m.
\end{equation}

Before the main results, we first give a lemma as follows:
\begin{lemma}
\label{lemma3}
Let $A,B,C\in \mathbf{R}^{n\times n}$ satisfy $AB=BC$ and $B^2=0$. Then the eigenvalue set of $A+B$ is the same as that of $A$.
\end{lemma}
\begin{pf}
Let rank$B=r$. When $r=0$ the assertion of the lemma is obviously right. When $r>0$, there is an $n$-by-$r$ matrix $M$ and a $r$-by-$n$ matrix $N$ such that
\begin{equation}
B=MN, \ \ \mathrm{rank}M=\mathrm{rank}N=r.
\end{equation}
Since $M$ has a full column rank, one can construct a nonsingular matrix $T=[M, T_1]$. Let $T^{-1}=[P^\mathrm{T}, Q^\mathrm{T}]^\mathrm{T}$. Then
\begin{equation}
\left[\begin{array}{cc}
       I_{r} & 0\\
       0 & I_{n-r}
     \end{array}\right]= T^{-1}T
     =\left[\begin{array}{cc}
       PM & PT_1\\
       QM & QT_1
     \end{array}\right],
\end{equation}
which implies that
\begin{equation}
\label{eq16}
QM=0, \ \ QB=QMN=0.
\end{equation}
Since $N$ has a full row rank and $BMN=B^2=0$, we have that $BM=0$. Thus
\begin{equation}
\label{eq17}
T^{-1}BT=\left[\begin{array}{c}
       P\\
       Q
     \end{array}\right]B[M\ \ T_1]=
     \left[\begin{array}{cc}
       0 & PBT_1\\
       0 & 0
     \end{array}\right].
\end{equation}
From $AMN=AB=BC$ and (\ref{eq16}), it follows that
\begin{equation}
\label{eq18}
QAM=QBCN^\mathrm{T}(NN^\mathrm{T})^{-1}=0.
\end{equation}
Thus
\begin{equation}
\label{eq19}
T^{-1}AT=\left[\begin{array}{c}
       P\\
       Q
     \end{array}\right]A[M\ \ T_1]=\left[\begin{array}{cc}
       PAM & PAT_1\\
       0 & QAT_1
     \end{array}\right].
\end{equation}
By (\ref{eq17}) and (\ref{eq19}), the proof is complete.
\hfill $\blacksquare$
\end{pf}

\begin{proposition}
\label{prop2}
Assume that the Laplacian matrix $L$ satisfies rank$(L)=m-1$ and the eigenvalues of $L$ are $\lambda_1=0$, $\lambda_2$, $\lambda_3$, $\dots$, $\lambda_m$. Then the characteristic polynomial of $\mathcal{T}(\cdot)$
is
\begin{equation}
\label{eq20}
s^m\prod_{i=2}^m(s-\lambda_i)\prod_{j=2}^m(s-2\lambda_j)\!\!\!\prod_{2\leq q<p\leq m}\!\!\!(s-\lambda_p-\lambda_q)^2.
\end{equation}

\end{proposition}
\begin{pf}
Since rank$(L)=m-1$ and $L\mathbf{1}_m=0$, there exists a nonsingular matrix $P=[\mathbf{1}_m, P_1]$ which results the Jordan canonical form of $L$ as follows:
\begin{equation}
\label{e10}
J\!=\!P^{-1}\!LP\!=\!\left[\!\begin{array}{cc}
                 0 &     \\
                  & \tilde J
                  \end{array}\!\right]
            \!\!=\!\!\left[\!\!\begin{array}{c|cccc}
                 0 &    \\ \hline
                  & \lambda_2& b_2 & &  \\
                  &     & \ddots & \ddots &  \\
                   &  &              & \ \lambda_{m-1}   &\ \  b_{m-1}\\
                  &  &        &                & \lambda_m
                  \end{array}\!\!\right],
\end{equation}
where every $\lambda_i$ is nonzero and each $b_i$ is one or zero.
It is easy to check that
\begin{eqnarray}
\label{e12}
&&(P^{-1}\otimes P^{-1})(L\otimes I_m+I_m\otimes L-\mathbf{1}_m\otimes \hat L)(P\otimes P)\nonumber \\
&=&J\otimes I_m+I_m\otimes J-\delta_1\otimes P^{-1}\hat L(P\otimes P)
 \nonumber \\
&=&J\otimes I_m+I_m\otimes J-\delta_1\otimes P^{-1}\hat L[\mathbf{1}_m\otimes P,\ \ P_1\otimes P]\nonumber \\
&=&J\otimes I_m+I_m\otimes J-\delta_1\otimes
[P^{-1}LP,\ \ P^{-1}\hat L(P_1\otimes P)]
\nonumber \\
&=&J\otimes I_m+I_m\otimes J-\delta_1\otimes
[J,\ \ P^{-1}\hat L(P_1\otimes P)]\nonumber \\
&=& \!\!\left[\!\!\begin{array}{cc}
                 0_{m\!\times \!m} &     \\
                  & \!\!\!\tilde J\otimes I_m
           \end{array}\!\!\right]\!\!+
                                \!\!\left[\!\begin{array}{cc}
                                 J &     \\
                                 & I_{m-1}\!\otimes \!J
                                \end{array}\!\right]\!\!-
                                          \!\!\left[\!\!\begin{array}{cc}
                                           J & \ \ P^{-1}\!\hat L(P_1\!\otimes \!P)    \\
                                              0    & 0
                                          \end{array}\!\!\right]\nonumber \\
&=& \!\left[\!\begin{array}{cc}
                 0_{m\times m} &    -P^{-1}\hat L(P_1\otimes P) \\
                   & \ \ \tilde J\otimes I_m+I_{m-1}\!\otimes J
           \end{array}\!\right].
  \end{eqnarray}
From (\ref{e12}), it follows that the characteristic polynomial of $L\otimes I_m+I_m\otimes L-\mathbf{1}_m\otimes \hat L$ is just (\ref{eq20}).
\par
In the following, we use Lemma \ref{lemma3} to complete the proof.
Let
\begin{eqnarray}
~~A&=&L\otimes I_m+\!I_m\otimes L-\mathbf{1}_m\otimes \hat L, \nonumber \\
 ~~B&=&\hat L\otimes \mathbf{1}_m=
                                     \left[\begin{array}{cccc}
                                    \mathbf{1}_mL_1 &    \\
                                          & \mathbf{1}_mL_2 &    \\
                                               &   & \ddots \\
                                                &  && \mathbf{1}_mL_m
                                                \end{array}\right] . \nonumber
\end{eqnarray}
From $L\mathbf{1}_m=0$, it follows that $B^2=0$, $(I_m\otimes L)B=0$ and $(\mathbf{1}_m\otimes \hat L)B=0.$ Now, we construct a matrix $C$ such that $(L\otimes I_m)B=BC$, which is equivalent to
\begin{equation}
\label{e19}
l_{ij} \mathbf{1}_mL_j= \mathbf{1}_mL_iC_{ij}, \ \ \forall \ 1\leq i,j\leq m.
\end{equation}
When $L_i=0$, we have $l_{ij}=0$, which means that $C_{ij}$ can be any $m$-by-$m$ matrix.
When $L_i\neq 0$, it is easy to check that $C_{ij}=l_{ij}L_i^\mathrm{T}(L_iL_i^\mathrm{T})^{-1}L_j$ satisfies (\ref{e19}). So, there exists a matrix $C$ such that $AB=BC$. Therefore, by Lemma \ref{lemma3}, $A-B$ and $A$ have the same characteristic polynomial. The proof is complete. \hfill $\blacksquare$
\end{pf}
By Proposition \ref{prop2}, we have obtained all the eigenvalues of the linear transformation $\mathcal{T}(\cdot)$. Now we can determine the exact exponential synchronization rate.
\begin{theorem}
Consider the high-dimensional Kuramoto model (\ref{eq3}) limited on  the unit sphere $\mathbf{S}^{n-1}$. If the interconnecting digraph with the adjacency matrix $(a_{ij})$ has a spanning tree, then the exponential synchronization rate is exactly
the smallest non-zero real
part of the eigenvalues of Laplacian matrix $L$.
\end{theorem}
\begin{pf}
Since the interconnecting digraph with the adjacency matrix $(a_{ij})$ has a spanning tree, the Laplacian matrix $L$ satisfies rank$L=m-1$ \cite{Ren2005Consensus}. So by Proposition \ref{prop2}, we get the characteristic polynomial of $\mathcal{T}(\cdot)$ as shown in (\ref{eq20}), where $\lambda_1=0$, $\lambda_2$, $\dots$, $\lambda_m$ are the eigenvalues of $L$.
By Lemma \ref{lemma2} and $\lambda_1=0$, we get the the characteristic polynomial of $T_{33}$ in (\ref{eq11}) as follows:
\begin{equation}
\label{eq24}
\prod_{i=2}^m(s-\lambda_i)\!\!\prod_{2\leq q<p\leq m}\!\!\!(s-\lambda_p-\lambda_q).
\end{equation}
From (\ref{eq11}), (\ref{eq20}) and (\ref{eq24}), it follows that the characteristic polynomial of $T_{11}$ is
\begin{equation}
\label{eq25}
\prod_{j=2}^m(s-2\lambda_j)\!\!\prod_{2\leq q<p\leq m}\!\!\!(s-\lambda_p-\lambda_q),
\end{equation}
Without loss of generality, we assume that $\lambda_2$ has the smallest nonzero real part of the eigenvalues of $L$. Therefore, by (\ref{eq25}) we see that the decay rate of $E(t)$ is $2\mathrm{Re}(\lambda_2)$, which implies that $|e_{ij}(t)|\leq c \mathrm{e}^{-2\mathrm{Re}(\lambda_2)t}$ for a constant $c>0$. Considering $|e_{ij}(t)|=e_{ij}(t)=\frac{1}{2}||r_i(t)-r_j(t)||^2$, we conclude that the exponential synchronization rate is $\mathrm{Re}(\lambda_2)$.\hfill $\blacksquare$
\end{pf}
\begin{remark}
The graph condition that the digraph has a directed spanning tree is
the weakest condition for synchronization. Actually, if the digraph has no a directed spanning tree, there exist at least two independent strongly connected components. Since  there is no any information interaction between two independent strongly connected components, it is impossible to achieve synchronization for all the possible initial states.
\end{remark}
\section{Conclusions}
For the high-dimensional Kuramoto model with identical oscillators and a general digraph admitting a spanning tree, the exponential synchronization rate has been accurately determined by using the matrix Riccati differential equation of the synchronization error dynamics. In our future work, we will try to generalize the main results and  method to non-identical Kuramoto oscillators and some other generalized high-dimensional Kuramoto models.

\bibliographystyle{abbrv}        
\bibliography{autosam}

\end{document}